# THREE-DIMENSIONAL (3D) TENSOR-BASED METHODOLOGY FOR CHARACTERIZING 3D ANISOTROPIC THERMAL CONDUCTIVITY TENSOR

**Dihui Wang[1], Heng Ban[1,*], Puqing Jiang[2,*]**

[1]Swanson school of engineering of University of Pittsburgh, Pittsburgh, US

[2]School of Energy and Power Engineering, Huazhong University of Science and Technology, Wuhan, Hubei 430074, China

*Corresponding authors: hengban@pitt.edu (H. B.); jpq2021@hust.edu.cn (P. J.)

## ABSTRACT

The increasing complexity of advanced materials with anisotropic thermal properties necessitates more generic and efficient methods to determine three-dimensional (3D) anisotropic thermal conductivity tensors with up to six independent components. Current methods rely on a vector-based framework that can handle only up to four independent components, often leading to inefficiencies and inaccuracies. We introduce Three-Dimensional Spatially Resolved Lock-In Micro-Thermography (3D SR-LIT), a novel optical thermal characterization technique combining a 3D tensor-based framework with an efficient area-detection experimental system. For simple tensors (e.g., x-cut quartz, $k_{xz} = k_{yz} = 0$), our method reduces uncertainty by over 50% compared to vector-based methods. For complex tensors with six independent components (e.g., AT-cut quartz), $2\sigma$ uncertainties remain below 12% for all components. A novel adaptive mapping approach enables high-throughput data acquisition (40 seconds to 3 minutes, depending on tensor complexity), over 35 times faster than current methods, and accommodates samples with 200 nm surface roughness. Extensive numerical validation on 1,000 arbitrary anisotropic tensors ranging from 1 to 1,000 $\mathrm{Wm^{-1}K^{-1}}$ further validates the robustness of this methodology. This work highlights significant advancements in thermal characterization of complex anisotropic materials.

## 1. INTRODUCTION

The development of advanced materials has accelerated rapidly, driven largely by initiatives like the Materials Project and Google DeepMind[1–4], which rely on extensive material property databases for data-driven discoveries[5–9]. However, a major deficiency in these databases is the lack of data on thermal transport properties, particularly for thermal conductivity. To date, only a few hundred out of the nearly 100,000 laboratory-synthesized materials in the Inorganic Crystal Structure Database have experimentally measured thermal conductivity valueS11[–12]. This shortfall is even more critical for materials exhibiting complex anisotropic thermal behavior, which are more challenging to characterize. The absence of a generalized and efficient method for determining anisotropic thermal properties hampers the progress of computational materials research and limits the practical applications of advanced materials in key fields such as electronics[13–17], thermal management[18,19], and photonics[20,21], where anisotropic properties offer unique advantages over isotropic materials.

The principal thermal conductivity tensor ($\boldsymbol{k}$) in a three-dimensional (3D) Cartesian coordinate system can have up to six independent components, given that in the absence of magnetic fields, $\boldsymbol{k}$ is symmetric due to the Onsager reciprocity relations[22,23]. An ideal thermal characterization technique should not only be capable of determining these six independent components but also offer high-throughput (HT) data acquisition and ease of experimentation.

Existing methods fall short of these criteria. The fundamental challenge lies in the reliance on vector-based methodologies, which project the second-order tensor into first-order tensors (vectors), and then attempt to reconstruct the full thermal conductivity tensor. This projection is typically achieved by creating an almost one-dimensional (1D) heating event, as seen in contact methods like the 3-ω methods[24–29] and non-contact approaches such as elliptical-beam time-domain thermoreflectance (TDTR)[30] and TDTR imaging[31]. Alternatively, the projection can also be achieved by direction-dependent acquisition, such as beam-offset frequency-domain thermoreflectance (BO-FDTR) [32] and BO-TDTR[33], two-dimensional time-resolved magneto-optical Kerr effect (TR-MOKE)[34] and spatial-domain thermoreflectance (SDTR)[35,36]. From a mathematical perspective, this projection and reconstruction process inherently results in a lossy interpretation. Consequently, while these techniques perform well for isotropic materials or those with simple anisotropy, they struggle to accurately measure more complex thermal conductivity tensors, particularly those with small off-diagonal components. To date, all studies have demonstrated the ability to determine at most one non-zero off-diagonal component [32,35,37,38]. Additionally, experiments built on vector framework inherently lacks efficiency as it requires serially probing[30,35,39].

Recent advancements in two-dimensional (2D) tensor-based analysis, specifically Spatially Resolved Lock-In Thermography (SR-LIT), have enabled efficient and accurate determination of in-plane thermal conductivity tensors. Unlike vector-based methods, SR-LIT utilizes the 2D spatially resolved phase response map to directly determine $\boldsymbol{k}_{\text{in}}$ without reducing it to a lower order[38]. However, SR-LIT is still limited to two-dimensional measurements and cannot capture the full 3D thermal conductivity tensor, which includes essential cross-plane components like $k_{xz}$, $k_{yz}$, and $k_{zz}$.

In this work, we present the Three-Dimensional Spatially Resolved Lock-In Micro-Thermography (3D SR-LIT), a novel thermal characterization technique based on a 3D tensor framework designed to efficiently and comprehensively determine the full anisotropic thermal conductivity tensor, including all six independent components. Our method employs an adaptive mapping approach that integrates phase and amplitude data from one to three orthogonal surfaces, depending on the complexity of the tensor, to directly determine the thermal conductivity tensor without the need for projection. For simple anisotropic tensors (e.g., tensors where $k_{xz} = k_{yz} = 0$), we demonstrate measurements on fused silica, sapphire, and x-cut quartz, showing that a single-surface measurement integrating phase and amplitude maps is sufficient, reducing result uncertainties by up to 60% compared to conventional vector-based methods. For complex anisotropic tensors, such as AT-cut quartz with six independent components, phase maps from three mutually perpendicular surfaces provide comprehensive sensitivity to all tensor components, maintaining $2\sigma$ uncertainty within 12% for each component. Instrumentally, we present a modified SR-LIT system featuring rapid data acquisition (40 seconds to 3 minutes, depending on tensor complexity), achieving a time efficiency over 35 times superior to traditional methods. Additionally, we demonstrate measurements of sample with surface roughness $r_a = 200$ nm (and up to 5 μm[38]), greatly expanding its applicability compared to conventional optical methods like TDTR or FDTR that requires mirror smooth surface with surface roughness smaller than 20 nm.

Through an extensive numerical study of 1000 arbitrary anisotropic tensors, with diagonal components ranging from 1~1000 $\text{Wm}^{-1}\text{K}^{-1}$, we demonstrate the robustness and accuracy of this technique. This tensor-based framework extends beyond a single technique and can be integrated with existing experiment methods, such thermoreflectance techniques and modulated photothermal radiometry[40,41]. Therefore, it represents a major improvement in efficiency and adaptability for characterizing advanced materials with complex anisotropy, facilitating both experimental research and practical applications.

## 2. METHOD

### 2.1 Instrumentation

The 3D SR-LIT platform is an all-optical setup with micrometer resolution, as shown in the schematic in Figure. 1(a). A continuous wave (CW) fiber-coupled laser (Thorlabs S4FC785) serves as the heating source, which is modulated directly by a function generator at a fixed frequency $f_0$. This modulated beam is focused onto the sample surface using a tube lens to create periodic heating, thereby inducing a spatially varying thermal response.

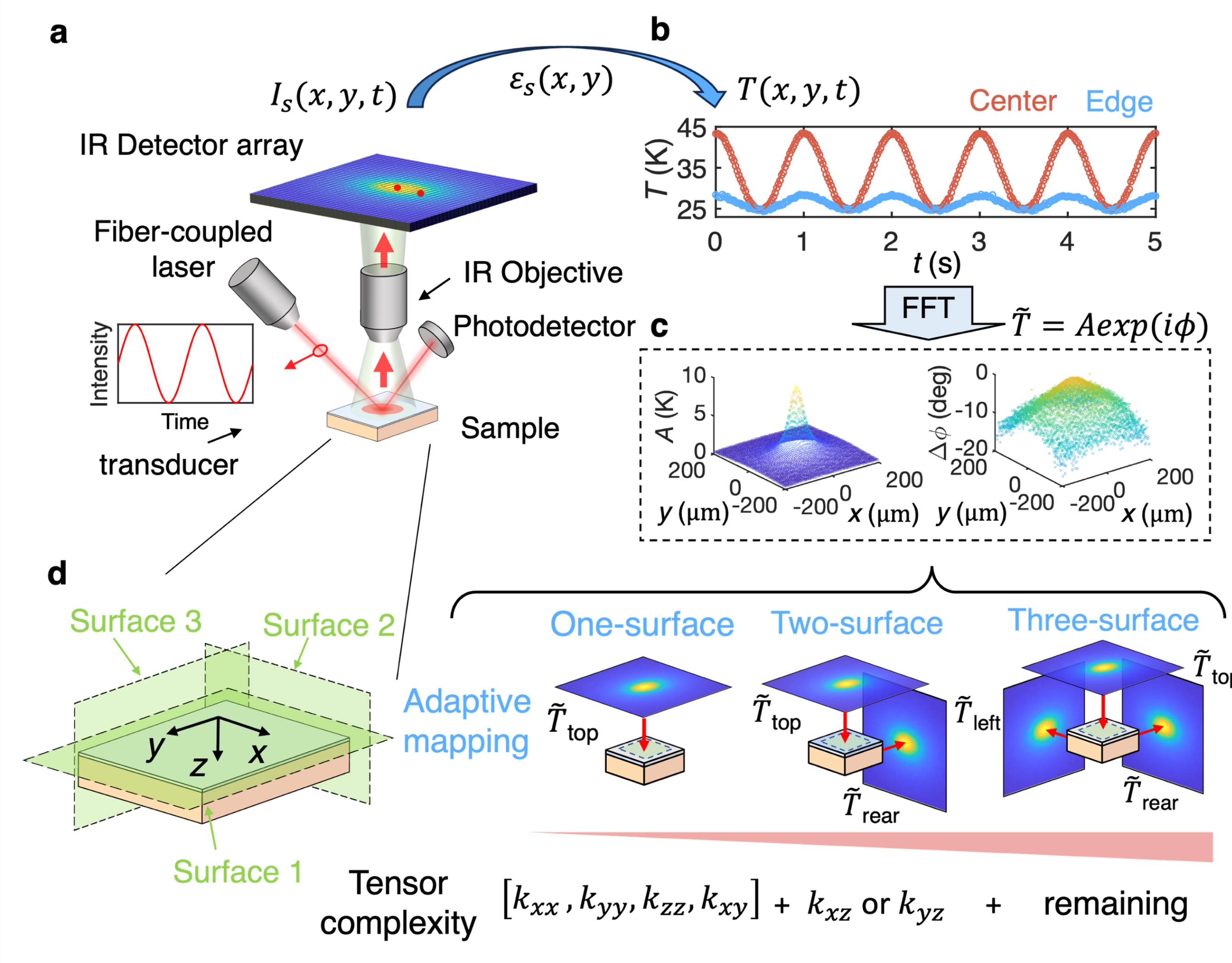

Figure. 1 (a) Schematic illustration of the 3D SR-LIT platform. (b) Overview of the 3D data acquisition process using FFT for signal extraction. (c) Example of a differential phase map $\Delta\phi(x, y)$ and an amplitude map $A(x, y)$, illustrating spatially resolved thermal response for data analysis. (d) Adaptive data mapping approach, with phase and amplitude maps collected from one to three surfaces, enabling accurate extraction of all components of the thermal conductivity tensor.

The reflected pump beam is captured by a photodetector (Thorlabs PM16-121 with PM100USB), with its intensity $P_r(t)$ recorded over time.

On the detection side, we utilize a mid-wavelength infrared (IR) micro-thermography system (QFI InfraScope MWIR Temperature Mapping Microscope), which captures the thermal radiation emitted from the sample in the wavelength range of 2.4 to 5 μm. The IR imaging system offers significant advantages over traditional single-point detection methods, such as those used in frequency-domain thermoreflectance (FDTR) or time-domain thermoreflectance (TDTR)[10,38], by enabling rapid acquisition of full-field thermal images. The infrared radiation from the sample is collected using a focal plane array (FPA) detector with a pixel size of 24 μm and a resolution of 500×500 pixels. With a 4× objective lens, the system achieves a refined pixel resolution of 6 μm, providing a field of view (FOV) of approximately 3000×3000 $\mu m^2$.

The surface temperature map, $T_s(x, y, t)$, is determined from the captured IR signal using the Stephan-Boltzmann law, which relates the emitted radiation to the temperature and the emissivity of the surface (see details in Supplementary S1). To ensure consistent absorption of the pump laser beam and efficient IR emission, all samples are coated with a 100-nm-thick Titanium (Ti) transducer layer. Before coating, the sample surface is cleaned via ion etching. The Ti layer is deposited using electron beam evaporation (Plassys Electron Beam Evaporator MEB550S), and its thickness is verified using a profilometer (Bruker DektaXT). The thermal conductivity of the Ti layer is estimated at approximately $15\ \mathrm{Wm^{-1}K^{-1}}$, calculated using the Wiedemann-Franz law based on four-point probe electrical resistivity measurements (Keysight B1500A with probe station).

During measurements, the IR detector captures the spatially resolved temperature response at a sampling rate of 53.4 Hz, enabling data acquisition in only 40 seconds per surface. The combination of high-resolution spatial mapping, accurate temperature detection, and adaptive phase/amplitude analysis provides significant advantages over conventional thermal characterization techniques, enabling reliable determination of the full 3D thermal conductivity tensor for a wide range of materials.

### 2.2 Data analysis

The 3D tensor analysis involves transforming the spatially resolved temperature response into a format that allows for accurately determining the entire thermal conductivity tensor. This process combines frequency-domain transformations, sensitivity mapping, and nonlinear regression to fit the experimental data to the theoretical thermal diffusion model.

### 2.2.1 Fast Fourier Transform and Signal Extraction

After collecting the raw temperature response data $T_s(x, y, t)$, as illustrated in Figure. 1(b), the time-domain signals are transformed into the frequency domain using a Fast Fourier Transform (FFT) (see details of FFT in Supplementary S2). This yields amplitude $A(x, y)$ and differential phase $\Delta\varphi(x, y)$ maps (Figure. 1(c)) for the sample, as well as a DC amplitude map. The amplitude and differential phase maps are crucial for thermal analysis, while the DC amplitude map provides information about the total temperature rise.

To ensure high signal quality, the data are filtered using a signal-to-noise ratio (SNR) threshold. Only data points with SNR values above a specified cutoff (set as 10 in this study) are used in subsequent analysis, eliminating regions with unreliable measurements. The resulting filtered maps retain critical spatial information while minimizing the influence of noise. Further details on SNR filtering can be found in Supplementary S3.

### 2.2.2 Adaptive Mapping for Tensor Extraction

The next step involves constructing an adaptive data map based on the complexity of the thermal conductivity tensor (Figure. 1(d)). For determine simple anisotropic tensors with $k_{xz} = k_{yz} = 0$, an integrated signal map including both phase and amplitude map measured from single-surface is sufficient. In these cases, the phase map provides sensitivity for extracting in-plane components $k_{xx}$ and $k_{yy}$, while the amplitude map provides sensitivity to the cross-plane component $k_{zz}$. For more complex anisotropic tensors with six independent components, phase maps from three orthogonal surfaces (i.e., the $x - y$, $x - z$, and $y - z$ planes) are utilized. As will be demonstrated in Section 3, this adaptive mapping approach ensures high sensitivity for each tensor component while optimizing measurement efficiency.

The amplitude map cannot be directly compared with the model, as the actual absorbed laser power, $P_{\text{absorbed}}$, is uncertain due to factors like variable absorptivity and optical attenuation. To address this challenge, we introduce a power coefficient $\beta$, which correlates $P_{\text{absorbed}}$ to the monitored power $P_r$ of the reflected pump beam from the sample surface. The power coefficient $\beta$ can be determined through calibration on a standard reference sample of isotropic fused silica. It is calculated through the relation $\beta A_{\text{sim}}(P_r) = A_{\text{measured}}$, where $A_{\text{sim}}(P_r)$ represents the simulated amplitude assuming a heating power of $P_r$, and $A_{\text{measured}}$ is the actual measured amplitude. Once calibrated, the same coefficient $\beta$ is applied to subsequent samples to correct for variations in power.

### 2.2.3 Nonlinear Regression and Model Fitting

Prior to fitting, all data maps are consolidated into one single data vector, which serves as input for a non-linear regression algorithm. The algorithm minimizes least-squares error through iterative adjustment of all desired tensor components, optimizing alignment between experimental data and simulated thermal response. The simulated thermal response is computed based on a 3D multilayer heat diffusion equation for each pixel in the adaptive map and considers various input parameters such as the thermal properties of the transducer layer (e.g., thermal conductivity, heat capacity), the substrate properties, and the thermal boundary conductance between the transducer and the substrate (see details of the model in Appendix S1).

It is crucial to characterize the spot size of the pump beam prior to fitting the thermal conductivity. The spot size is determined directly from the normalized amplitude map (the amplitude normalized by the maximum amplitude at the heating center, $A_{\mathrm{norm}} = A/A_{\mathrm{max}}$)[35,38]. This approach eliminates the need for additional spot characterization experiments, such as knife-edge measurements, and enhances the reliability of the measurement by mitigating unintentional spot size variations across different measurement sets (see validation in Supplementary S4). Other input parameters, such as the thermal conductivity of the metal transducer, heat capacity of the metal transducer and substrates, as well as the thermal boundary conductance between the transducer and substrate, are obtained from literatures (see detailed values in Supplementary S5, Table S1). It should be noted that the signal exhibits negligible sensitivity to the thermal boundary conductance in all cases, thus eliminating the need for an accurate value.

## 3. RESULTS

### 3.1 Validation on isotropic thermal conductivity tensor

We first determine the power coefficient $\beta$ using a standard isotropic fused silica sample. The modulation frequency of the pump laser was set to 5 Hz to achieve an in-plane thermal diffusion length $d_r = \sqrt{k_r/\pi f_0 C}$ that exceeds three times the laser spot size, thereby maximizing sensitivity to in-plane thermal transport. This frequency was selected to remain below the camera's Nyquist frequency of 27 Hz, ensuring precise phase and amplitude capture.

Figure. 2(a) and 2(b) present the measured differential phase ($\Delta\phi$) and amplitude ($A$) map, with filtering applied to enhance fitting quality (filtering protocol detailed in Supplementary S6). After filtering, the $\Delta\phi$ map retains a ring-shaped profile, and the $A$ map preserves a centralized response region. The phase data alone provides sufficient information to accurately determine the isotropic thermal conductivity tensor, given its sensitivity to both $k_{xx}$ and $k_{yy}$, with minimum sensitivity to $k_{zz}$ across the entire region, as shown in Figure. 2(d)-2(f) (sensitivity formalism provided in Appendix. A2). For isotropic materials, $k_{zz}$ was computed as the geometric mean $k_{zz} = \sqrt{k_{xx}k_{yy}}$. The measured thermal conductivities for fused silica were $k_{xx}^{\pm 2\sigma} = 1.46 \pm 0.05\ \mathrm{Wm^{-1}K^{-1}}$, $k_{yy}^{\pm 2\sigma} = 1.48 \pm 0.05\ \mathrm{Wm^{-1}K^{-1}}$, and $k_{zz}^{\pm 2\sigma} = 1.47 \pm 0.05\ \mathrm{Wm^{-1}K^{-1}}$, using a literature value of $C_v = 1.62\ \mathrm{Jcm^{-3}K^{-1}}$[44] for the volumetric heat capacity as input (see uncertainty formalism in Appendix. A3). These values are consistent with both our previous results[35] and those found in the literatures[44]. With the thermal properties of the fused silica sample established, we determined the power coefficient $\beta$ by fitting the amplitude map. This yields $\beta = 0.67$, which was subsequently applied to correct the input power for further measurements.

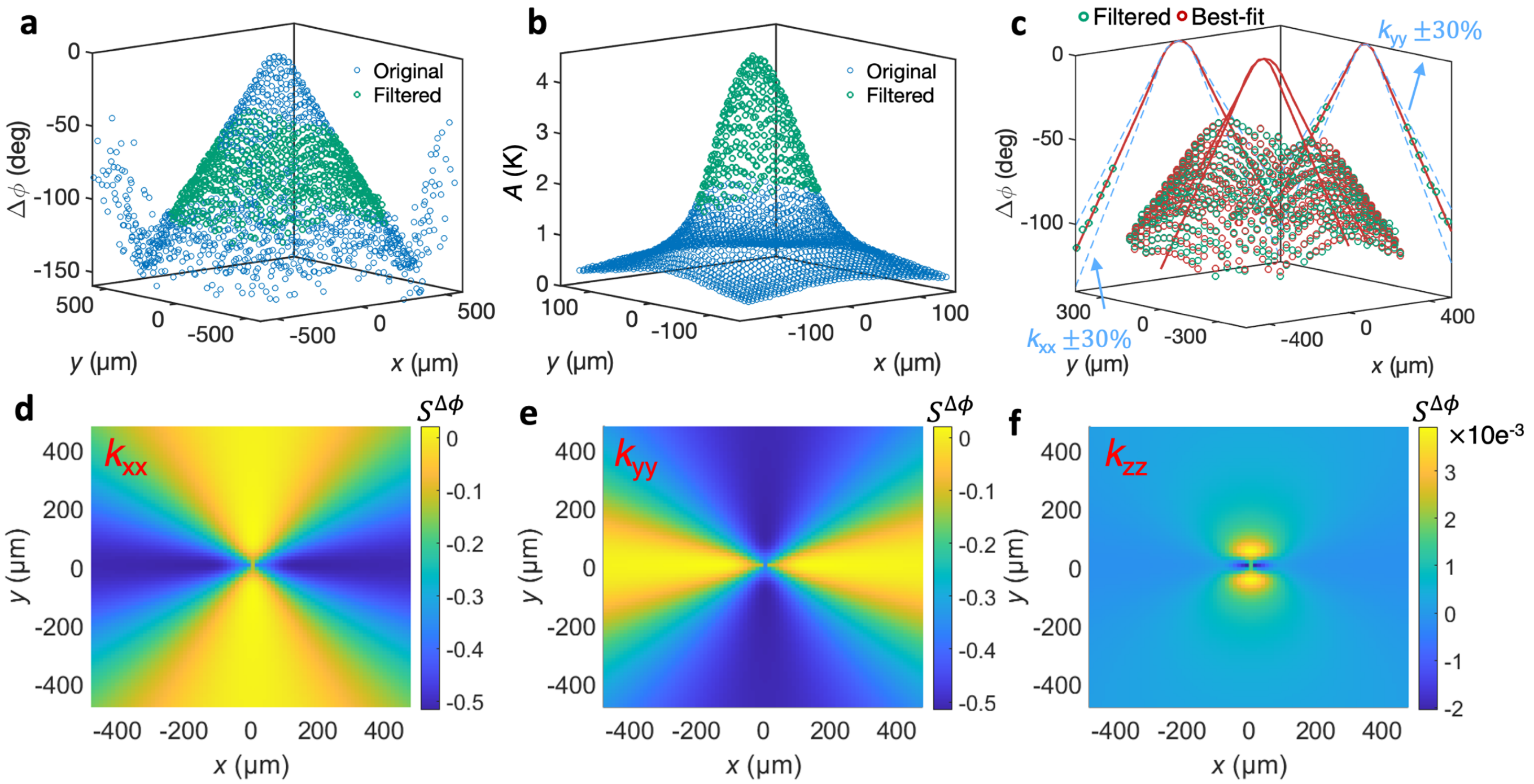

Figure. 2 Experimental phase and amplitude maps for fused silica: (**a**) Measured $\Delta\phi$ map (blue open circle) with filtered data (orange open circle) to improve fitting accuracy; (**b**) Measured amplitude map following the same filtering process; (**c**) Comparison of the filtered $\Delta\phi$ map (green dots) and the best-fit $\Delta\phi$ map. Data measured along the x- and y-axes are projected on the y- and x-plane, respectively. Blue dashed curves indicating 30% bounds of the best-fitted tensor element along the offset direction ($k_{xx}$ for x-axis data and $k_{yy}$ for y-axis data) are included as guides of sensitivity. (**d-f**) Sensitivity maps of $\Delta\phi$ with respect to in-plane ($k_{xx}$, $k_{yy}$) and cross-plane ($k_{zz}$) thermal conductivity, demonstrating how changes in thermal conductivity influence the phase signals near the heating center.

We then apply the method on a (0001)-oriented sapphire sample. While sapphire is often regarded as isotropic in thermal conductivity, with negligible anisotropy reported in the literature[32,45,46], we treated $k_{xx}$, $k_{yy}$, and $k_{zz}$ as three independent unknowns. The adaptive map, incorporating both the $\Delta\phi$ map and the $A$ map, was analyzed to extract thermal properties, yielding $k_{xx}^{\pm 2\sigma} = 37.1 \pm 2.5\ \mathrm{Wm^{-1}K^{-1}}$, $k_{yy}^{\pm 2\sigma} = 37.1 \pm 2.5\ \mathrm{Wm^{-1}K^{-1}}$, and $k_{zz}^{\pm 2\sigma} = 38.5 \pm 3.1\ \mathrm{Wm^{-1}K^{-1}}$, align well with the literature values (summarized in Supplementary S5, Table S2). A similar measurement on $(11\bar{2}0)$-oriented sapphire produced consistent results, further validating the isotropy assumption. The slightly higher uncertainties observed for sapphire, compared to fused silica, were primarily due to reduced signal-to-noise ratios when measuring high thermal conductivity materials under limited laser power. This limitation could be addressed in future setups by employing higher-power lasers.

### 3.2 Validation on simple anisotropic thermal conductivity tensors

To evaluate the applicability of our method to simple anisotropic thermal conductivity tensors, we demonstrate measurement on a $(11\bar{2}0)$-oriented x-cut quartz. The principal axis of the sample is intentionally rotated by $20°$, yielding a target tensor with four independent parameters $k_{xx}, k_{yy}, k_{zz}$, and $k_{xy}$ and negligible cross-plane off-diagonal components ($k_{xz} = k_{yz} = 0$) (see Figure. 3(i)). This type of tensor still poses challenges for traditional methods that require multi-directional measurements that typically involve manual reconfiguration of optics (add citations). By contrast, our 3D tensor approach utilizes the same procedure used for isotropic cases, streamlining the process and enhancing efficiency.

The sensitivity map of the phase for in-plane components ($k_{xx}$, $k_{yy}$, and $k_{xy}$) reveals distinct patterns, enabling simultaneous determination (Figure. 3, top row). Notably, unlike $k_{xx}$ and $k_{yy}$, which primarily affect the gradient of the phase along corresponding axis[47,48], the off-diagonal component $k_{xy}$ induces a rotational effect in the phase map around the *z* axis. This effect is illustrated in Figure. 3(j), which compares phase maps of an isotropic tensor ($k = 10\,\mathrm{Wm^{-1}K^{-1}}$) and an anisotropic tensor with $k_{xy} = 3\,\mathrm{Wm^{-1}K^{-1}}$. However, as shown in Figure. 3(d), the $\Delta\phi$ map exhibits near-zero sensitivity to $k_{zz}$ (on the order of $10^{-4}$) across the entire region, indicating that $k_{zz}$ cannot be determined from $\Delta\phi$.

The amplitude map provides the additional information needed to determine $k_{zz}$. As shown in the second row of Figure. 3, the sensitivity of the amplitude map to $k_{zz}$ is significant, particularly near the heating center, where its absolute value reaches 0.5, which is 5 to 10 times higher than the sensitivity to the in-plane components. To aid those familiar with vector-based sensitivity analysis, we also plotted the sensitivities of $\Delta\phi$ and $A$ as functions of the angle $\gamma$ relative to the x-axis (see Figure. 3(k) and 3(l)). The red dashed circles in the first and second row represent selected points at a fixed radius, where $d_{\Delta\phi} = 6r(\theta)$ for $\Delta\phi$ and $d_{\mathrm{A}} = 0.5r(\theta)$ for $A$. This representation clearly demonstrates the distinct trends for each tensor component, with sensitivity differences large enough to meet the requirement for independent determination. Therefore, we combine the phase map and the amplitude map as our integrated signal.

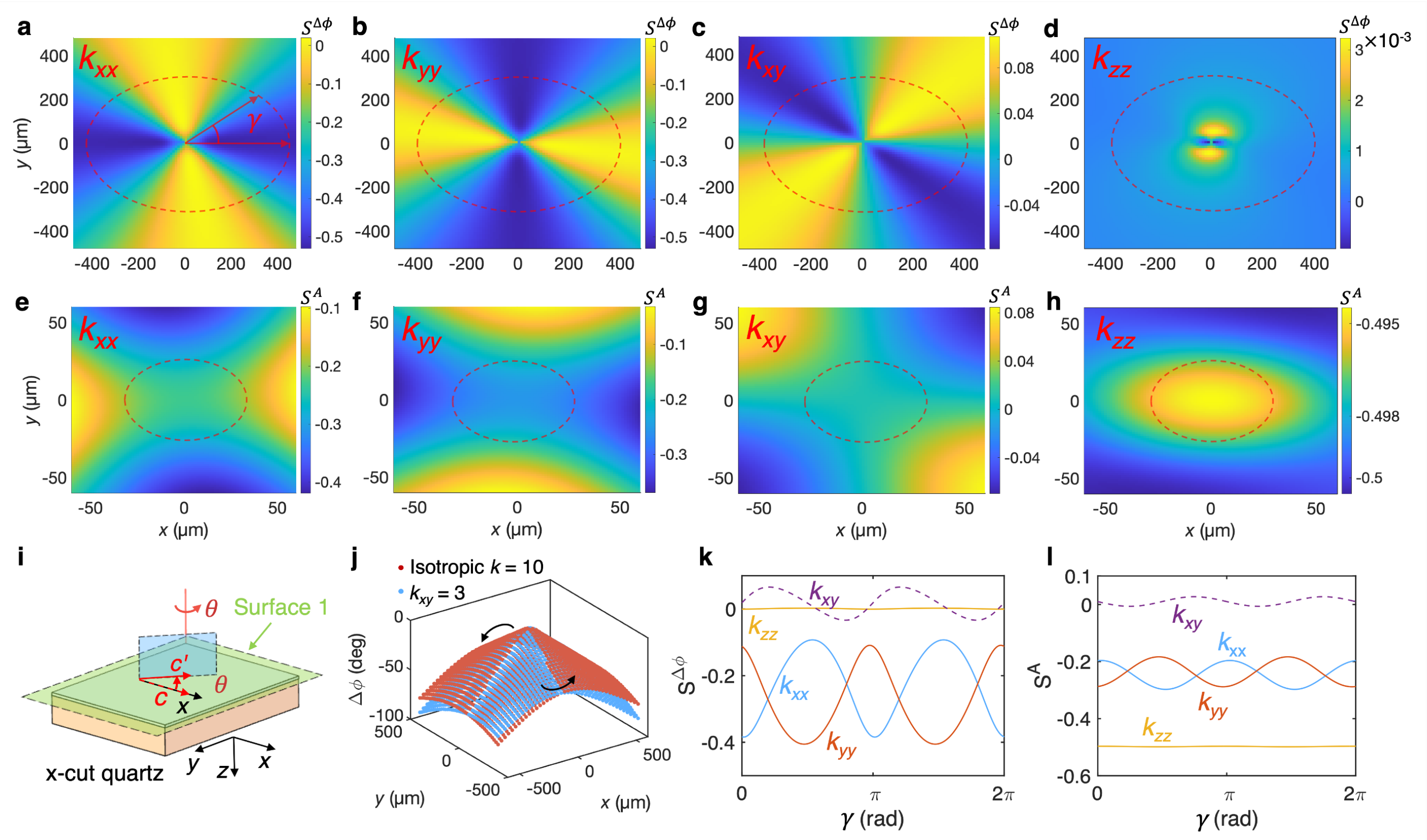

Figure. 3 (**a-d**) sensitivity maps of differential phase $\Delta\phi$ to in-plane components ($k_{xx}$, $k_{yy}$, $k_{xy}$) and cross-plane component ($k_{zz}$). (**e-f**) sensitivity maps of amplitude $A$ to in-plane components ($k_{xx}$, $k_{yy}$, $k_{xy}$) and cross-plane component ($k_{zz}$). The red dashed circles represent selected points at a fixed radius, where $d_{\Delta\phi} = 6r(\theta)$ for $\Delta\phi$ and $d_{\mathrm{A}} = 0.5r(\theta)$ for $A$. (**i**) Illustration of the x-cut quartz sample rotated by an angle $\theta$, with $c'$ representing the new position of the c-axis. The transparent green plane highlights the measurement Surface 1, while the transparent blue plane indicates the plane of the c-axis after rotation. (**j**) Comparison of $\Delta\phi$ maps for isotropic and anisotropic cases, showing how the off-diagonal component ($k_{xy}$) introduces rotation in the phase map around the $z$ axis. (**k-l**) Sensitivity plots of the $\Delta\phi$ and $A$ signals to in-plane components ($k_{xx}$, $k_{yy}$, $k_{xy}$) and cross-plane component ($k_{zz}$) as a function of angle $\gamma$ about the $x$-axis demonstrate the distinct trends for each tensor component, enabling their independent determination.

The final best-fit thermal conductivity tensor for the rotated x-cut quartz is:

$$\boldsymbol{k}_{\text{rotated}}^{\pm 2\sigma} = \begin{bmatrix} 11.4 \pm 0.53 & -1.62 \pm 0.11 & 0 \\ -1.62 \pm 0.11 & 7.4 \pm 0.43 & 0 \\ 0 & 0 & 7.1 \pm 0.45 \end{bmatrix} (\text{Wm}^{-1}\text{K}^{-1}).$$

For comparison, we examined the fitting quality of a vector-based analysis, which incorporates phase data from four directions (0° , 45°, 90°, and 135°) along with a single amplitude measurement at the heating center. This approach yielded:

$$\boldsymbol{k}_{\text{vector}}^{\pm 2\sigma} = \begin{bmatrix} 11.8 \pm 0.9 & -1.47 \pm 0.32 & 0 \\ -1.47 \pm 0.32 & 7.1 \pm 0.75 & 0 \\ 0 & 0 & 7.2 \pm 0.76 \end{bmatrix} (\text{Wm}^{-1}\text{K}^{-1}).$$

The tensor-based analysis demonstrates superior accuracy and substantially reduces uncertainty compared to the vector-based approach. Specifically, uncertainties in the diagonal components are reduced by 50%, while uncertainties in the off-diagonal components decrease by approximately 60%.

Using the derived $\boldsymbol{k}_{\text{rotated}}$, the rotation angle $\theta_{\text{r}}$ was deduced as $\theta_r = \frac{1}{2}\arctan\left(\frac{2k_{xy}}{k_{xx}-k_{yy}}\right)$[49,50], resulting in $\theta_r = 18.9° \pm 1.3°$. Finally, the principal thermal conductivity tensor of x-cut quartz can be retrieved by applying rotation matrix and yields (see definition of rotation matrix in Supplementary S7):

$$\boldsymbol{k}_{\text{principal}}^{\pm 2\sigma} = \begin{bmatrix} 11.9 \pm 0.5 & 0 & 0 \\ 0 & 6.9 \pm 0.3 & 0 \\ 0 & 0 & 7.1 \pm 0.4 \end{bmatrix} (\text{Wm}^{-1}\text{K}^{-1}).$$

This result tensor exhibits excellent agreement with measurements on the same sample without rotation and with literature values, where $k_{xx,\text{lit}} = 12\ \text{Wm}^{-1}\text{K}^{-1}$ and $k_{yy,\text{lit}} = k_{\text{zz,lit}} = 6.8\ \text{Wm}^{-1}\text{K}^{-1}$ [44].

### 3.3 Validation on complex cross-plane anisotropy

The subsequent experimental validation examines complex cross-plane anisotropy, where either $k_{xz}$ or $k_{yz}$ is non-zero. We demonstrate this using a case with non-zero $k_{xz}$, noting that the procedure applies equivalently for non-zero $k_{yz}$ is identical. The tested sample is AT-cut quartz, a type of quartz crystal sliced at an angle of approximately $\psi = 35°$ relative to its principal *c*-axis (see Figure. 4(a)). With the crystal orientation known, we align the plane containing the *c*-axis parallel to the $xz$-plane of the designed coordinate system (Figure. 4(b)). Under this configuration, the target anisotropic tensor can be computed based on values measured from x-cut quartz: $\boldsymbol{k}_{\text{AT,target}} = \begin{bmatrix} 10.4 & 0 & -2.3 \\ 0 & 7 & 0 \\ -2.3 & 0 & 8.6 \end{bmatrix}$ ($\text{Wm}^{-1}\text{K}^{-1}$). Interestingly, sensitivity analysis reveals that $\Delta\phi$ exhibits increased sensitivity to $k_{zz}$ in tensors where $k_{xz} \neq 0$, compared to cases with $k_{xz} = 0$. This phenomenon can be explained using Fourier's law, which describes the in-plane heat flux $q_x$:

$$q_x = -[k_{xx}\ k_{xy}\ k_{xz}]\begin{bmatrix}\partial T/\partial x\\ \partial T/\partial y\\ \partial T/\partial z\end{bmatrix}.$$

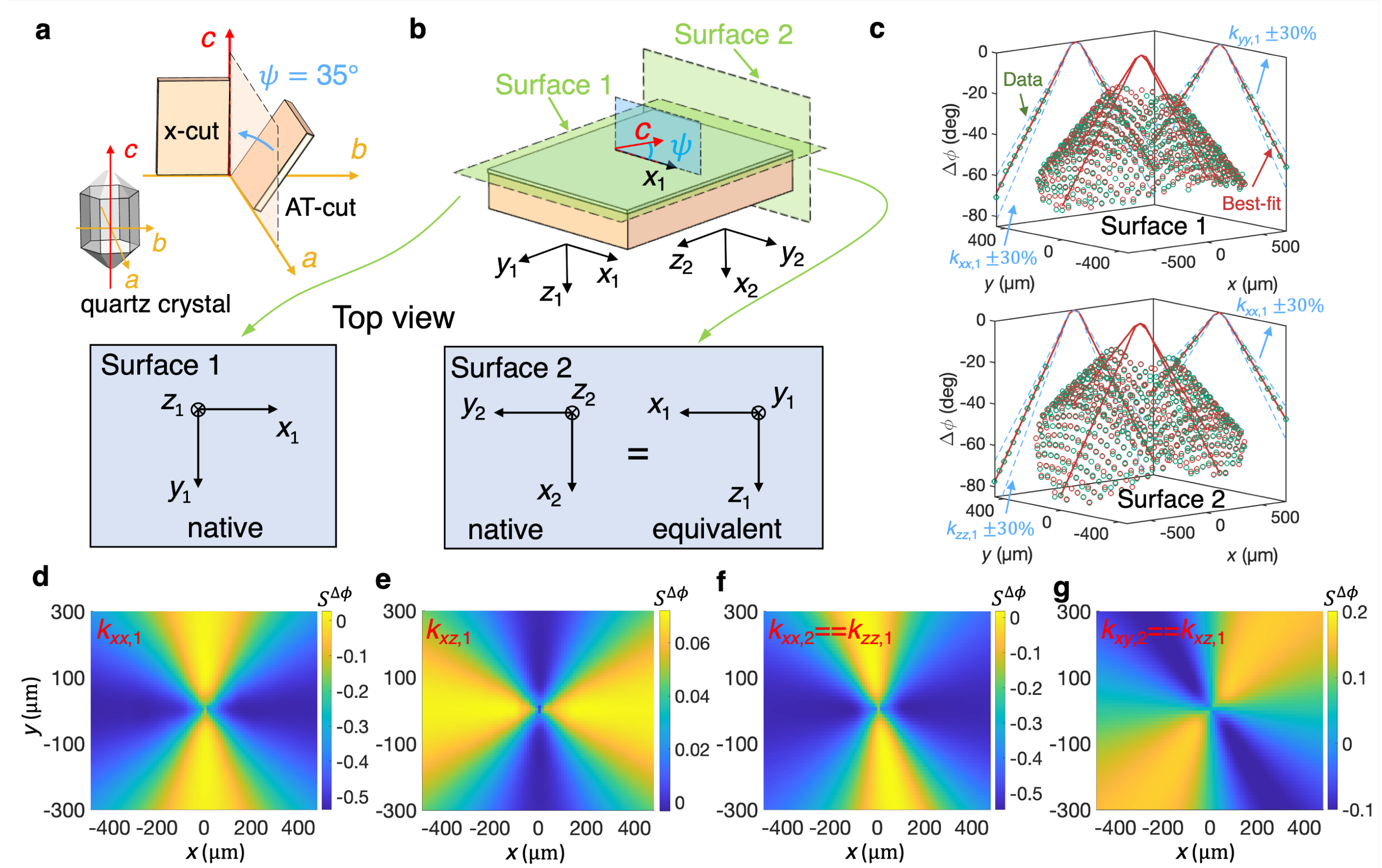

Figure. 4 (**a**) Schematic illustration of the x-cut and AT-cut quartz crystal orientations, with the angle $\psi$ indicating the AT-cut angle relative to the c-axis. (**b**) Schematic configuration of aligned AT cut quartz with the two-surface measurement. The transparent blue plane indicates the plane of the plane c-axis, while two transparent green planes highlight the two measurement surfaces. Two coordinate systems are defined based on observations at Surface 1 and Surface 2, with subscripts indicating the respective observation planes. The second row provides the top views of Surface 1 (left) and Surface 2 (right). Each top view shows both their native coordinate system, and the equivalent coordinate system based on surface 1. (**c**) Measured differential phase signals for Surface 1 (top) and Surface 2 (bottom), with blue dashed lines indicating sensitivity to diagonal tensor components. (**d-g**) Sensitivity maps for phase signals on both surfaces, demonstrating distinct sensitivity patterns for the diagonal and off-diagonal components ($k_{xx}$, $k_{xy}$, $k_{zz}$, and $k_{xz}$). This two-surface measurement approach ensures accurate determination of the full anisotropic thermal conductivity tensor for AT-cut quartz.

When $k_{xz} \neq 0$, the $\partial T/\partial z$ term contributes to the in-plane heat flux $q_x$, indirectly linking $k_{zz}$, which governs the cross-plane heat transport, to in-plane behavior. However, this interaction remains subtle, and $k_{zz}$ cannot be reliably determined solely from $\Delta\phi$.

Here we focus on the characteristic of the off-diagonal term $k_{xz}$. From Figure. 4(d) and 4(e), we observe that the sensitivity pattern of the phase to $k_{xz}$ is similar to that of $k_{xx}$, but with a significantly smaller magnitude and opposite sign, making it difficult to separate $k_{xz}$ from $k_{xx}$. To accurately determine $k_{xz}$, we propose conducting the same experiment on Surface 2, as defined in Figure. 4(b).

Based on the rotation matrix and coordinate system arrangement shown in Figure. 4(b), the native tensor at Surface 2 can be expressed in terms of the native tensor at Surface 1:

$$\boldsymbol{k_2} = \begin{bmatrix} k_{xx,2} & k_{xy,2} & k_{xz,2} \\ k_{xy,2} & k_{yy,2} & k_{yz,2} \\ k_{xz,2} & k_{yz,2} & k_{zz,2} \end{bmatrix} = \begin{bmatrix} k_{zz,1} & k_{xz,1} & k_{yz,1} \\ k_{xz,1} & k_{xx,1} & k_{xy,1} \\ k_{yz,1} & k_{xy,1} & k_{yy,1} \end{bmatrix} = \begin{bmatrix} 8.6 & -2.3 & 0 \\ -2.3 & 10.4 & 0 \\ 0 & 0 & 7 \end{bmatrix} (\mathrm{Wm^{-1}K^{-1}}).$$

We observe that $\boldsymbol{k}_2$ satisfies the definition of simple anisotropy, and thus can be determined based on the integrated phase map and amplitude map as previous section. However, we propose to use an alternative approach. Notably, the two non-zero cross-plane components at surface 1, $k_{xz,1}$ and $k_{zz,1}$, are transformed into in-plane components at surface 2, with $k_{xz,1} = k_{xy,2}$ and $k_{zz,1} = k_{xx,2}$. The $\Delta\phi$ shows high sensitivity to both parameters (see Figure. 4(f) and 4(g)). Collectively, we construct an adaptive signal map using only the $\Delta\phi$ maps measured at two surfaces, and the amplitude signal, which provides sensitivity to $k_{zz,1}$, can be omitted. This approach not only provides distinct and high sensitivity for each desired tensor component but also eliminates the need for a reference sample measurement, streamlining the overall procedure.

A 2 mm slice is cut from a 5 mm thick AT-cut quartz sample to serve as the second sample, with the 5 mm side oriented upward to provide Surface 2 for measurement (see the design in Supplementary S8.1). After cutting, the roughness of Surface 2 is determined using a stylus profilometer (Bruker DektaXT surface profiler) and measured as $r_a = 200$ nm. It is worth noting that, SR-LIT accommodates measurements on sample with surface roughness up to 5 μm when using only the phase map[38]. This capability arises from the intrinsic advantage of IR detection and the large characteristic length of the heating event (see details in Supplementary S8.2). Therefore the 200 nm roughness is well within the workable range. The smallest lateral dimension of 5 mm is verified through COMSOL Multiphysics® finite element modeling (FEM) analysis and confirmed to satisfy the semi-infinite condition (see Supplementary S9). The surface is coated with a metallic transducer, following the same treatment as Surface 1.

During the data analysis, both $\Delta\phi$ measured at Surface 1 and Surface 2 are fed into the model, with four unknowns: $k_{xx,1}$, $k_{yy,1}$, $k_{zz,1}$, and $k_{xz,1}$. When modeling Surface 2, the native tensor properties are substituted with the equivalent properties at Surface 1: $k_{xx,2} = k_{zz,1}$, $k_{yy,2} = k_{xx,1}$, $k_{zz,2} = k_{yy,1}$, and $k_{xy,2} = k_{xz,1}$. Care must also be taken to ensure that the raw coordinate system of the data map at Surface 2 is properly rotated to align with the native coordinate system at Surface 2 (Figure. 4(b)), as the equivalent properties no longer holds if this alignment is not achieved. The resulting thermal conductivity tensor is:

$$\boldsymbol{k}_{\mathrm{AT}}^{\pm 2\sigma} = \begin{bmatrix} 10.1 \pm 0.5 & 0 & -2.2 \pm 0.15 \\ 0 & 7.2 \pm 0.4 & 0 \\ -2.2 \pm 0.15 & 0 & 8.4 \pm 0.4 \end{bmatrix} (\mathrm{Wm^{-1}K^{-1}}),$$

which exhibits good agreement with the target tensor.

### 3.4 Validation of 3D tensor analysis on anisotropic tensor with six independent terms

The final experimental validation addresses the most complex case: a 3D anisotropic tensor with six independent terms. To create a 3D anisotropic tensor, the AT-cut quartz is rotated by $45°$ (see Figure. 5(a)). Consequently, the native thermal conductivity tensor at Surface 1 is:

$$\boldsymbol{k}_{\mathrm{AT,rotated,1}} = \begin{bmatrix} 8.7 & -1.68 & -1.66 \\ -1.68 & 8.7 & 1.66 \\ -1.66 & 1.66 & 8.6 \end{bmatrix} (\mathrm{Wm^{-1}K^{-1}}).$$

In the previous section, we achieved high sensitivity to the off-diagonal terms $k_{xy,1}$ and $k_{xz,1}$ by acquiring signals from Surface 1 and Surface 2, respectively. Building on this, we introduce a measurement at a third surface, which is orthogonal to Surface 1 and Surface 2. The corresponding tensor at Surface 3 satisfies:

$$\boldsymbol{k_3} = \begin{bmatrix} k_{xx,3} & k_{xy,3} & k_{xz,3} \\ k_{xy,3} & k_{yy,3} & k_{yz,3} \\ k_{xz,3} & k_{yz,3} & k_{zz,3} \end{bmatrix} = \begin{bmatrix} k_{yy,1} & k_{yz,1} & k_{xy,1} \\ k_{yz,1} & k_{zz,1} & k_{xz,1} \\ k_{xy,1} & k_{xz,1} & k_{xx,1} \end{bmatrix},$$

where the cross-plane property $k_{yz,1}$ is transformed into an in-plane property ($k_{xy,3} = k_{yz,1}$) in Surface 3 (Figure. 5(a)). Subsequently, the adaptive signal is constructed using $\Delta\phi$ maps from all three surfaces, providing excellent and distinct sensitivity patterns to all six tensor components (detailed sensitivity maps are provided in Supplementary S10).

Two angled slices are cut from the AT-cut quartz sample, with known crystal orientation, at angles of 45° and -45°, respectively, creating Surface 2 and Surface 3 for the measurement (see Figure. 5(e) and 5(f)). The $\Delta\phi$ maps measured at these three surfaces are fed into the model, which considers six unknowns expressed as terms in the tensor at Surface 1: $[k_{xx,1}, k_{yy,1}, k_{zz,1}, k_{xy,1}, k_{xz,1}, k_{yz,1}]$. The best-fit maps are shown in Figure. 5(b)-(d), and the resulting thermal conductivity tensor is:

$$\boldsymbol{k}^{\pm 2\sigma}_{\text{AT,rotated}} = \begin{bmatrix} 8.9 \pm 0.5 & -1.6 \pm 0.2 & -1.73 \pm 0.21 \\ -1.6 \pm 0.2 & 8.4 \pm 0.5 & 1.64 \pm 0.18 \\ -1.73 \pm 0.21 & 1.64 \pm 0.18 & 8.5 \pm 0.5 \end{bmatrix} (\text{Wm}^{-1}\text{K}^{-1}),$$

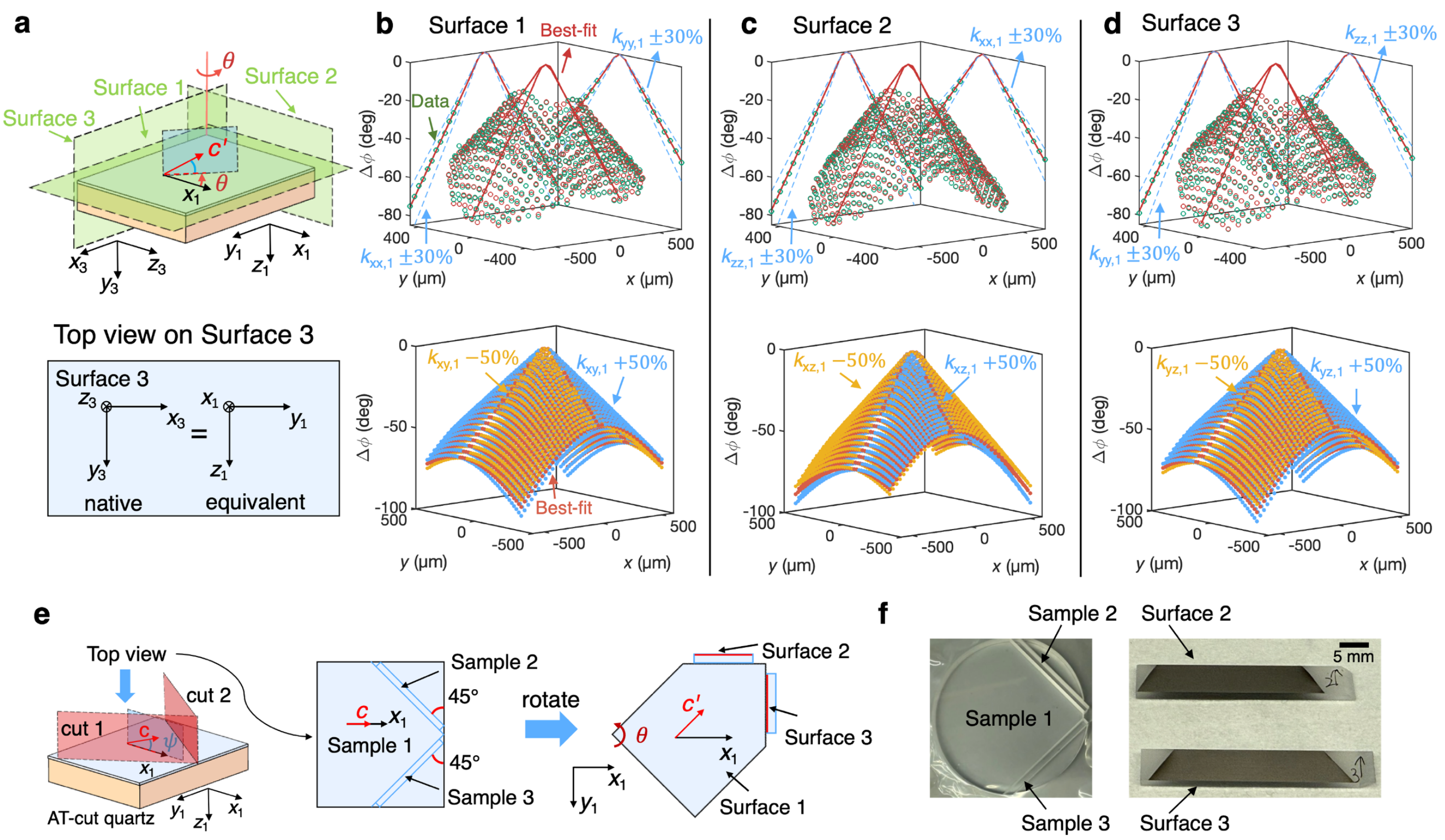

Figure. 5 (**a**) Schematic of the three-surface measurement setup for AT-cut quartz, rotated by $\theta$, illustrating the relative orientations of the three surfaces used for tensor analysis. For Surface 3, the native tensor is expressed in an equivalent form using components oriented as on Surface 1. (**b-d**) Best-fit differential phase maps for the three surfaces, with measured data (green) and best-fit model (red) overlaid. Blue dashed lines indicate sensitivity to diagonal components of the thermal conductivity tensor. For each surface, the components for guide are chosen differently to highlight each surface sensitive to different components. The second row shows sensitivity guides for off diagonal terms. (**e**) The illustration of sample preparation for the three-surface measurement, including two cuts at $\pm 45°$ relative to the $c$-axis of the AT-cut quartz. (**f**) Images of samples. This multi-surface approach ensures high accuracy and precision in determining all six independent components of the 3D anisotropic thermal conductivity tensor.

which shows good agreement with literature values, with well-controlled uncertainties. For each surface, sensitivity guidelines are plotted using different diagonal components to demonstrate the unique sensitivity of each surface. In the second row of Figure. 5(b)-5(d), we also present the sensitivity boundaries of ±50% for different off-diagonal components, confirming that each off-diagonal component exhibits distinct sensitivity trends, enabling simultaneous determination.

### 3.5 Numerical validation

In the final section, we validate the 3D tensor-based analysis through numerical simulations across diverse anisotropy scenarios. The simulated thermal signals incorporate realistic experimental conditions: (1) a 100 nm titanium transducer layer, (2) random noise levels matching x-cut quartz, and (3) an elliptical heating pattern with $w_x = 60\ \mu\text{m}$ and $w_y = 40\ \mu\text{m}$. The validation begins with two extreme cases and is followed by a comprehensive study involving 1000 randomly generated anisotropic tensors.

**Case 1: Extreme diagonal anisotropy**

The first tensor, $\boldsymbol{k}_{\text{extrme},1} = \begin{bmatrix} 3000 & 20 & 0 \\ 20 & 100 & 0 \\ 0 & 0 & 10 \end{bmatrix}$ ($\text{Wm}^{-1}\text{K}^{-1}$), features extreme diagonal anisotropy ($k_{xx}/k_{yy}$=30 and $k_{xx}/k_{zz} = 300$) and significant in-plane off-diagonal anisotropy ($\tau_{xy} = \frac{\sqrt{k_{xx}k_{yy}}}{k_{xy}} = 27$). Given $k_{xz} = k_{yz} = 0$, we compare single-surface amplitude-phase mapping and three-surface phase mapping methods. The one-surface mapping yields:

$$\boldsymbol{k}_{\text{1surf},1}^{2\sigma\%} = \begin{bmatrix} 2956 \pm 4.8\% & 21.4 \pm 36.4\% & 0 \\ 21.4 \pm 36.4\% & 102.3 \pm 4.4\% & 0 \\ 0 & 0 & 10.4 \pm 8.9\% \end{bmatrix} (\text{Wm}^{-1}\text{K}^{-1}),$$

while three-surface mapping gives:

$$\boldsymbol{k}_{\text{3surf},1}^{2\sigma\%} = \begin{bmatrix} 3039 \pm 4.41\% & 21.8 \pm 46.1\% & 0 \\ 21.8 \pm 46.1\% & 98.7 \pm 4.2\% & 0 \\ 0 & 0 & 10.2 \pm 4\% \end{bmatrix} (\text{Wm}^{-1}\text{K}^{-1}).$$

Both methods demonstrate high accuracy. The single-surface achieves 21% lower uncertainty for $k_{xz}$, while the three-surface achieves 55% lower uncertainty for $k_{\text{zz}}$. Considering its experimental simplicity, we recommend the single-surface method for tensors with simple anisotropy.

**Case 2: Large off-diagonal anisotropy**

The second tensor, $\boldsymbol{k}_{\text{extrme},2} = \begin{bmatrix} 300 & 9 & -6 \\ 9 & 150 & 6 \\ -6 & 6 & 250 \end{bmatrix}$ $(\mathrm{Wm^{-1}K^{-1}})$, exhibits significant off-diagonal anisotropy with $\tau_{xy} = 24, \tau_{xz} = 46$, and $\tau_{yz} = 32$. Using three-surface mapping, the results are:

$$\boldsymbol{k}_{\text{3surf},2}^{2\sigma\%} = \begin{bmatrix} 304.2 \pm 4.2\% & 8.32 \pm 29.7\% & -5.1 \pm 38.2\% \\ 8.32 \pm 29.7\% & 148.7 \pm 4.3\% & 6.65 \pm 66.5\% \\ -5.1 \pm 66.5\% & 6.65 \pm 38.2\% & 248.5 \pm 4.2\% \end{bmatrix} (\mathrm{Wm^{-1}K^{-1}}).$$

All components exhibit good accuracy, although the off-diagonal components exhibit higher uncertainties. Nevertheless, compared to a vector-based method using twelve directional measurements (0°, 45°, 90°, and 135° on three surfaces), the 3D tensor analysis reduces the uncertainties of off-diagonal terms by 59%.

**General numerical study**

Next, we demonstrate a generic numerical study across over 1000 tensors with random anisotropy. The diagonal components $k_{xx}$, $k_{yy}$, and $k_{zz}$ are randomly chosen between $1$ and $1000\ \mathrm{Wm^{-1}K^{-1}}$, while the off-diagonal components satisfy $k_{ij}^2 < k_{ii}k_{jj}$ (where $i, j = x, y, z$), ensuring the thermal conductivity tensor remains positive-definite.

The performance was evaluated based on two key metrics: accuracy and precision. For accuracy, the best-fit results for $k_{xx}$ and $k_{xz}$, as shown in Figure. 6(a) and 6(b), demonstrated excellent performance across all cases. To provide a more comprehensive assessment, we employed a violin plot to visualize the distribution of the percentage differences ($e_{\text{diff}} = \frac{(k_{\text{designed}} - k_{\text{fitted}})}{k_{\text{designed}}} \times 100\%$) for all tensor components. As illustrated in Figure. 6(c), 98% of the components exhibited percentage errors below 6%. The elevated error in the remaining 2% primarily happens when fitting tensors with significant off-diagonal anisotropy ($\tau_{ij}$) and the regression algorithm prematurely stopped due to iteration limit. By modifying the fitting process - extending the iteration limit and using intermediate values from the point of interruption - these tensors with large $\tau_{ij}$ successfully converge to errors below 6%.

While extending the iteration limit improved accuracy, it also increased regression time, which varied depending on tensor anisotropy and the choice of initial values. For nearly isotropic tensors, regression can be completed in about one minute, whereas for tensors with significant off-diagonal anisotropy $\tau_{ij}$ and poorly selected initial values, regression times can reach up to five hours. This limitation, however, could be mitigated by employing more powerful computational resources or machine learning models[51].

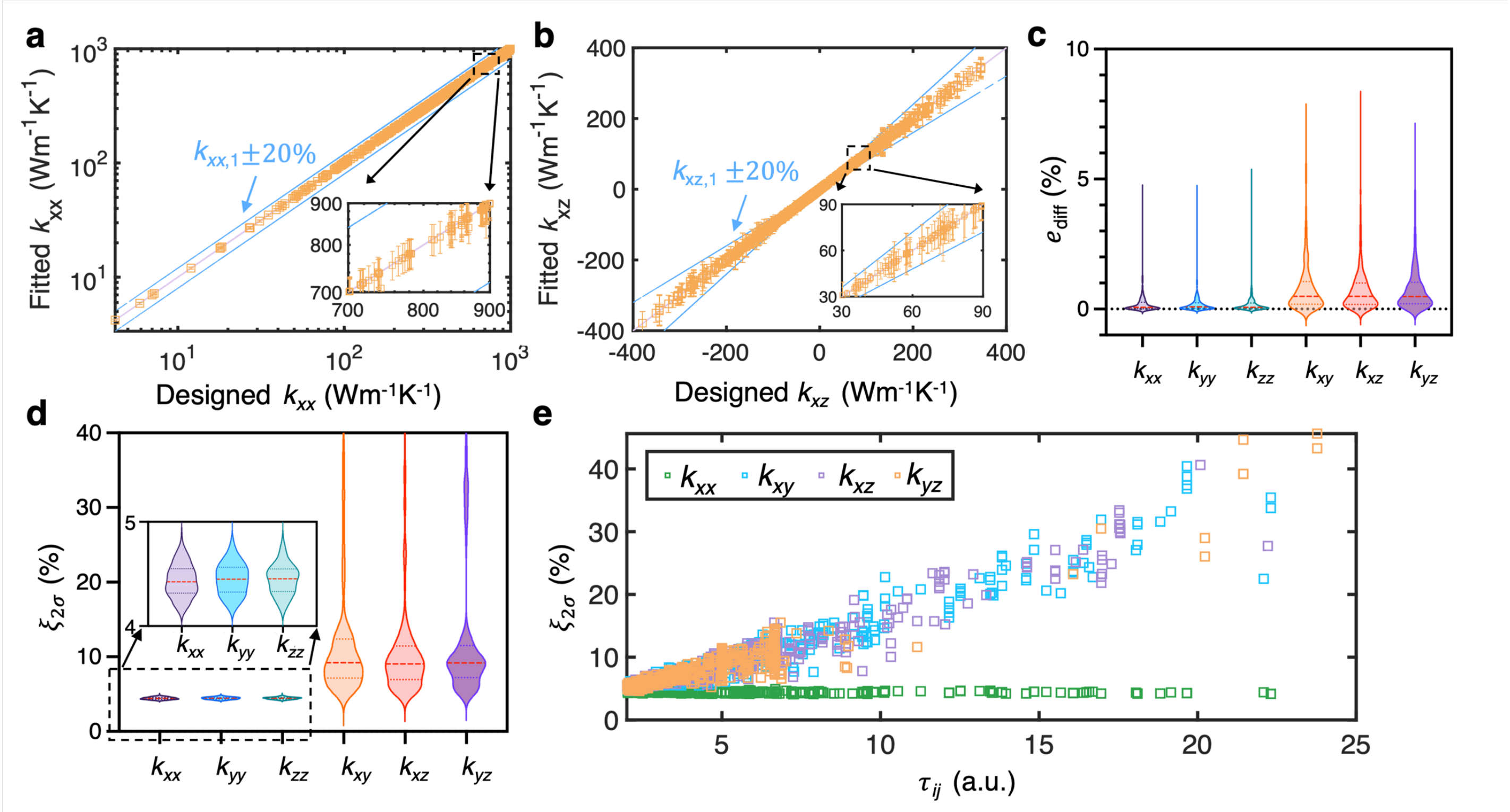

Figure. 6 Summary of numerical validation for the 3D tensor analysis method using over 1000 simulated anisotropic thermal conductivity tensors. (**a-b**) Best-fit results for in-plane ($k_{xx}$) and cross-plane ($k_{xz}$) thermal conductivity components, showing excellent accuracy compared to the designed values. The best-fit results are accompanied by $\pm 2\sigma$ uncertainty, with blue dashed lines indicating sensitivity by perturbating corresponding term by 20%. Insets provide a clearer view of the dense data regions. (**c**) Violin plot of percentage error for each tensor component. (**d**) Violin plot of percentage $2\sigma$ uncertainty $\xi_{2\sigma}$ with quartiles. (**e**) $\xi_{2\sigma}$ of $k_{xx}$, $k_{xy}$, $k_{xz}$, and $k_{yz}$, as a function of corresponding off-diagonal anisotropy $\tau_{ij}$.

Regarding precision, uncertainty was carefully analyzed. Figure. 6(a) and 6(b) show that the uncertainty for $k_{xx}$ and $k_{xz}$ were generally low, with most cases falling within $\pm 20\%$ under the $\pm 2\sigma$ guideline. Figure. 6(d) revealed that diagonal components ($k_{xx}$, $k_{yy}$, and $k_{zz}$) exhibited excellent precision with uncertainties below 6%, while off-diagonal terms showed slightly higher but acceptable uncertainties, with the third quartile around 13%.

We also explored the relationship between off-diagonal anisotropy ($\tau_{ij}$) and $\xi_{2\sigma}$, as this is crucial for predicting the precision of target tensors. Our analysis revealed that while the precision of diagonal terms was not affected by $\tau_{ij}$, the off-diagonal components exhibit a linear relation with it (Figure. 6(e)). For tensors with $\tau_{ij} < 17$, the 3D tensor method provided acceptable precision, with $\xi_{2\sigma}$ below 30%. It is worth noting that $\tau_{ij} = 17$ represents an exceptionally large value. For example, in reported cases such as 20°-rotated x-cut quartz, AT-cut quartz, and $\beta$-$Ga_2O_3$, the off-diagonal anisotropy typically remains below 8, where the expected $\xi_{2\sigma}$ using 3D tensor analysis is under 10%.

## 4. SUMMARY AND OUTLOOK

We introduce a novel 3D tensor-based methodology for thermal property characterization that overcomes the limitations of traditional vector-based methods, enabling generic and high-throughput determination of arbitrary 3D thermal conductivity tensors with enhanced accuracy and precision. Using demonstrative measurements, we illustrate the construction of adaptive 3D data maps, comprising phase /and amplitude distributions obtained from one or multiple surface constructions, tailored to the complexity of the thermal conductivity tensor. We demonstrate that for fused silica, sapphire, and x-cut quartz, a single-surface phase-amplitude map reduces uncertainties by up to 60% compared to vector-based methods. For complex tensors like AT-cut quartz, phase maps from three orthogonal surfaces ensure excellent sensitivities to all components, keeping $2\sigma$ uncertainties within 12%. Instrumentally, we propose a modified Spatially Resolved Lock-in Thermography (3D SR-LIT). This system features surface temperature mapping at micrometer resolution and enables rapid data acquisition in under three minutes. Finally, we demonstrate an extensive numerical study of 1000 arbitrary anisotropic tensors, with diagonal components ranging from 1~1000 $Wm^{-1}K^{-1}$, confirms the robustness of this technique.

Looking forward, incorporating a machine learning (ML) model could further improve the efficiency of the regression process[51]. Moreover, the impact of our study extends beyond the development of a single technique. This tensor-based framework can be seamlessly integrated with current experiment methods, such as thermoreflectance methods, fluorescence thermometry[52–54], and photoacoustic thermometry[55–57], among others. These features collectively render 3D tensor-based methodology a valuable approach for broad research and real-world applications, significantly improving the characterization of advanced materials with arbitrary anisotropy.

## SUPPLEMENTARY MATERIAL

See the Supplementary Material for extensive additional information about the experimental method and data analysis.

## DATA AVAILABILITY

The data that support the findings of this study are available from the corresponding authors upon reasonable request.

## ACKNOWLEDGMENT

D.W. acknowledges the financial support from the GSR of Swanson School of Engineering at the University of Pittsburgh. P.J. acknowledges the support of the National Natural Science Foundation of China N(NSFC) through Grant No. 52376058.

## CREDIT AUTHOR STATEMENT

**D.W.**: Conceptualization (lead); Experiment (lead); Formal analysis (lead); Validation (equal); Writing – original draft (lead); Writing – review & editing (equal). **P.J.**: Conceptualization (equal); Experiment (supporting); Validation (equal); Writing – review & editing (lead); **H.B.**: Conceptualization (equal); Funding acquisition (lead); Project administration (lead); Resources (lead); Writing – review & editing (equal).

## APPENDIX

### 1. Mathematical model

The 3D heat diffusion equation in cartesian coordinates with no heat-generation term is given by

$$C_v \frac{\partial T}{\partial t} = k_{xx}\frac{\partial^2 T}{\partial x^2} + k_{yy}\frac{\partial^2 T}{\partial y^2} + k_{zz}\frac{\partial^2 T}{\partial z^2} + 2k_{xy}\frac{\partial^2 T}{\partial x \partial y} + 2k_{xz}\frac{\partial^2 T}{\partial x \partial z} + 2k_{yz}\frac{\partial^2 T}{\partial y \partial z}, \tag{1}$$

where $C_v$ is the volumetric heat capacity, and $k_{ij}$ with different subscripts are corresponding elements in the thermal conductivity tensor $\boldsymbol{k}$,

$$\boldsymbol{k} = \begin{bmatrix} k_{xx} & k_{xy} & k_{xz} \\ k_{yx} & k_{yy} & k_{yz} \\ k_{zx} & k_{zy} & k_{zz} \end{bmatrix}. \tag{2}$$

The solution for solving the multi-layer 3D heat diffusion model has been well established in the literature[30,58] and can be found in Supplementary S11. Generally, equation (1) can be transformed into an ordinary differential equation (ODE) by applying the Fourier transform to time $t$ and both in-plane coordinates $x, y$, $T(x, y, z, t) \rightarrow$ $\Theta(u, v, z, \omega)$. The ODE of the multilayered system can then be solved by the thermal quadrupole approach. The resulting surface temperature is

$$\Theta_{\text{top}}(x, y, \omega) = \int_{-\infty}^{\infty} \int_{-\infty}^{\infty} \hat{G}(u, v, \omega) Q_0(u, v, \omega) e^{i2\pi(ux+vy)} dudv, \tag{3}$$

where $Q_0(u, v, \omega)$ is the intensity of absorbed heat flux after the Fourier transform in spatial and time domains, and $\hat{G}(u, v, \omega)$ is the Green's function. Since the heating spot radius and thermal penetration depth are significantly larger than the optical absorption depth, it is reasonable to assume surface absorption[59].

The next step is to determine the thermal response at each pixel. Assume the center of the probe pixel is located at $(x_c, y_c)$. Then, the weight function of the probe pixel can be written as

$$I_{\text{probe}}(x_c, y_c) = \int_{x_c-\frac{l_p}{2}}^{x_c+\frac{l_p}{2}} \int_{y_c-\frac{l_p}{2}}^{y_c+\frac{l_p}{2}} \frac{1}{l_p^2} dxdy, \tag{4}$$

where $l_p$ is the side length of the pixel. Finally, the probed thermal response for a pixel at $(x_c, y_c)$ is given by the weighted average of $\tilde{T}_{\text{top}}(x, y)$ by $I_{\text{probe}}(x_c, y_c)$:

$$H(x_c, y_c, \omega) = \frac{1}{l_p^2} \int_{x_c-\frac{l_p}{2}}^{x_c+\frac{l_p}{2}} \int_{y_c-\frac{l_p}{2}}^{y_c+\frac{l_p}{2}} \Theta_{top}(x, y, \omega) dxdy. \tag{5}$$

The thermal response $H(x_c, y_c, \omega)$ has a linear relation with measured data and will be evaluated numerically.

## 2. Sensitivity analysis

Sensitivity analysis is a valuable tool for evaluating the impact of different parameters on the signal and can guide the optimized experimental configuration. Here we use the sensitivity coefficient definition proposed by Gundrum *et al*[60],

$$S^{\gamma} = \frac{\partial \ln(\gamma)}{\partial \ln(\alpha)} \tag{6}$$

where $\gamma$ denotes the choice of signal and $\alpha$ denotes the parameter of interest. Here all input parameters are tested, including metallic transducer properties $k_m, C_m$, and transducer thickness $h_m$; substrate properties $k_{xx}$, $k_{yy}$, $k_{zz}$, $k_{xy}$, $k_{xz}$, $k_{yz}$, and $C_s$; and other inputs including interface thermal conductance between the metallic transducer and the substrate $G$, beam spot radius $w_x, w_y$, heating laser power $A_0$, and pixel side length $l_p$. With a low modulation frequency and small pixel size, we found that all the signals show negligible sensitivity to the film thermal property, $G$, $l_m$, and $l_p$, as such they are omitted in further analysis.

## 3. Uncertainty formalism

Our uncertainty formalism follows the work of Yang *et al.*[43] and Seber[42]. This full-error propagation formalism is based on the framework of regression analysis and can be applied to measurements where a known model is fit to one or more observable parameters using a least square algorithm. The loss function of the fitting is defined as:

$$R = \sum_{i=1}^{N} \left[ y_{d_{ofs,i}} - g\left(X_U, X_p, d_{ofs,i}\right) \right]^2, \quad (7)$$

where $y_i$ is the measured signal at the $i$-th offset spot, $g$ is the corresponding value evaluated by the thermal model, $X_U$ is the vector of unknown parameters and $X_p$ is the vector of input parameters. The best-fit unknown parameter will have uncertainty from both experimental noise and uncertainty of input parameters. For 3D anisotropic tensor, we consider six unknown parameters: $X_U = \left[k_{xx}, k_{yy}, k_{zz}, k_{xy}, k_{xz}, k_{yz}\right]^T$, and ten input parameters: $X_{pi} = \left[A_0, k_m, C_m, C_s, h_m, l_p, G_s, w_x, w_y\right]^T$. For those input parameters, their standard deviation is given in Supplementary S5 Table.S1. The resulting variances of unknown parameters are given in the format of variance-covariance matrix:

$$Var[X_U] = \begin{bmatrix} \sigma_{u_1}^2 & cov(u_1, u_2) & \dots \\ cov(u_2, u_1) & \sigma_{u_2}^2 & \cdots \\ \vdots & \vdots & \ddots \end{bmatrix}, \quad (8)$$

where the elements on the principal diagonal are the variance of the unknown parameters, and the value of $\pm 2\sigma$ (95% confidence interval) is used as the uncertainty of corresponding unknown parameters. Detailed derivation can be found in Supplementary S12.